\documentclass[10pt,a4paper,twoside]{article}

\input{spp.dat}

\begin{document}

\title{\TitleFont The Illusion of Rationality: Tacit Bias and Strategic Dominance in Frontier LLM Negotiation Games}

\author[1,2*]{Manuel S. Ríos\authorsep} 
\author[3]{Ruben F. Manrique\authorsep}
\author[4]{Nicanor Quijano\authorsep}
\author[1]{Luis F. Giraldo\lastauthorsep} 

\affil[1]{Department of Biomedical Engineering, Universidad de los Andes, Bogotá, Colombia}
\affil[2]{Center of Excellence in Analytics, Artificial Intelligence, and Information Governance, Bancolombia, Colombia}
\affil[3]{Department of Systems and Computing Engineering, Universidad de los Andes, Bogotá, Colombia}
\affil[4]{Department of Electrical and Electronics Engineering, Universidad de los Andes, Bogotá, Colombia}

\affil[*]{\corremail{ms.rios10@uniandes.edu.co}}




\begin{abstract}
\noindent
Large language models (LLMs) are increasingly being deployed as autonomous agents on behalf of institutions and individuals in economic, political, and social settings that involve negotiation. Yet this trend carries significant risks if their strategic behavior is not well understood. In this work, we revisit the NegotiationArena framework and run controlled simulation experiments on a diverse set of frontier LLMs across three multi-turn bargaining games: Buyer–Seller, Multi-turn Ultimatum, and Resource Exchange. We ask whether improved general reasoning capabilities lead to rational, unbiased, and convergent negotiation strategies. Our results challenge this assumption. We find that models diverge into distinct, model-specific strategic equilibria rather than converging to a unified optimal behavior. Moreover, strong numerical and semantic anchoring effects persist: initial offers are highly predictive of final agreements, and models consistently generate biased proposals by collapsing diverse internal valuations into rigid, generic price points. More concerningly, we observe dominance patterns in which some models systematically achieve higher payoffs than their counterparts. These findings underscore an urgent need to develop mechanisms to mitigate these issues before deploying such systems in real-world scenarios.


\end{abstract}

\maketitle
\thispagestyle{empty}  
\pagestyle{empty}

\section{Introduction}

Negotiation, a fundamental pillar in economic, political, and social interactions, is increasingly being delegated to autonomous artificial intelligence (AI) systems \cite{mastercard_agent_pay} \cite{gartner_agentic_ai_2025} \cite{hadfield2025economy}. 
Indeed, recent advances in the reasoning, planning, and communication capabilities of Large Language Models (LLMs) have encouraged many institutions and individuals to deploy LLM-based systems to strategically act on their behalf in negotiation scenarios \cite{park2023generative} \cite{peng2025diplomacyagent}. 
Consequently, substantial research efforts have emerged to benchmark and understand the actual capabilities of these systems in this domain \cite{bianchi2024well} \cite{kwon2024llms}.
However, the rapid pace of progress in the field forces us to continuously challenge the results of these initial works, as they become outdated with the release of more sophisticated AI systems.
This work addresses this gap by evaluating frontier models in negotiation scenarios and analyzing how progress in the field has impacted the strategic decision-making of these systems.

Understanding LLMs' strategic behavior is a critical question for economic stability and social fairness. The wide adoption of this technology will drive unprecedented interactions between AI systems and humans, as well as among AI systems themselves. These emergent multi-agent environments will present novel challenges for ensuring fairness and equality \cite{sharp2025agentic}. 
For instance, recent works provide empirical evidence suggesting that smaller models can be consistently exploited by larger and more complex models \cite{zhu2025automated}.
Moreover, an increasingly large body of research indicates that LLMs exhibit a variety of cognitive biases \cite{bianchi2024well} \cite{cheung2025large} \cite{lou2024anchoring} \cite{singh2024large}, which further exacerbate these vulnerabilities. This combination of exploitable cognitive biases and large capability imbalances creates the conditions for predatory economic dynamics.

As the reasoning capabilities of LLMs improve, as evidenced by their performance on numerous general benchmarks \cite{google2025gemini25} \cite{anthropic2025claude45}, one might naively expect, following classical game-theoretic intuitions \cite{nash1951non} \cite{schelling1960strategy}, that greater cognitive capabilities would induce convergence toward stable equilibrium strategies under fixed incentives. Generally, in these negotiation settings, such equilibrium behavior is commonly associated with rational and fair outcomes.
However, if this naive assumption is wrong, as supported by the empirical evidence presented in this work, then simply improving models' general capabilities is insufficient to guarantee rational and fair negotiation outcomes. This highlights the necessity of research agendas aimed at explicitly counteracting these strategic vulnerabilities and cognitive biases, moving beyond sole reliance on scaling.

Previous foundational work, such as NegotiationArena \cite{bianchi2024well}, provided some of the first insights into the systematic flaws of LLMs when confronting negotiation scenarios. Parallel research has expanded this evaluation to complex cooperative environments, demonstrating the potential and limitations of LLM-augmented agents to collaborate in benchmarks like Melting Pot \cite{smith2025evaluating} \cite{mosquera2025can}. Moreover, Zhu et al. \cite{zhu2025automated} empirically showed that, in AI-mediated customer markets, less capable AI agents can be systematically exploited by stronger counterparts. In this work, we revisit the NegotiationArena framework not only to provide updated results using frontier models, but also to show that these issues are not self-correcting with scale. Rather, frontier models occupy distinct, model-specific strategic equilibria in which reasoning flaws and cognitive biases remain clearly present.

In this work, we make three primary contributions. 
First, we show that systematic cognitive biases persist in the negotiation games we analyze. We identify strong numeric and semantic anchoring effects in frontier models. 
Second, we empirically challenge the naive assumption of convergence: in contrast to classical intuition, frontier models diverge into distinct, model-specific strategic equilibria, each developing its own unique “strategic signature.” 
Finally, we show that this divergence is not benign. By analyzing pairwise interactions across multiple games, we find that some state-of-the-art models can consistently dominate and systematically exploit weaker counterparts, providing empirical support for concerns about predatory dynamics in real-world deployments.

\section{Related Work}

\subsection{LLMs as Autonomous Agents and Multi-Agent Systems}

LLMs have evolved from passive text generators to agentic systems that interact with their environments. This transformation has been enabled by techniques that better elicit the reasoning and planning capabilities of base LLMs. In 2022, Wei et al. \cite{wei2022chain} proposed Chain-of-Thought (CoT), a seminal technique that improves LLMs' reasoning capabilities by prompting them to generate a sequence of intermediate steps before providing their final answer. This simple yet effective idea inspired subsequent work in which LLMs are prompted to verbalize their thought process before acting \cite{yao2022react} or to iteratively deliberate over a set of candidate thoughts before choosing a course of action \cite{yao2023tree}. 

Building on these advances, researchers have deployed LLMs in simulated scenarios that require long-horizon planning to achieve their goals. For instance, Wang et al. introduced Voyager \cite{wang2023voyager}, an LLM-powered agent capable of continuously exploring and acquiring new skills within the open-ended world of Minecraft. More recent work has explored the emergent dynamics of LLM-based multi-agent systems. Park et al. \cite{park2023generative} deploy “generative agents” in an interactive sandbox environment designed to approximate human social behavior; these agents can remember, reflect, and plan, giving rise to complex emergent social patterns. Recent studies have further formalized these dynamics by proposing metrics to measure the cooperative resilience of such agents when facing environmental disruptions \cite{chacon2025cooperative}.

This rapid progress has naturally motivated interest in moving beyond purely game-like environments, towards domains where LLMs have direct real-world consequences. In particular, it has inspired the deployment of LLM-based agents in economic and political negotiation settings \cite{mastercard_agent_pay}. However, these deployments demand rigorous evaluation to ensure that the impressive performance gains observed on benchmarks and in simulated environments actually translate into robust, fair, and strategically sound behavior in high-stakes negotiation scenarios.

\subsection{LLMs in Negotiation and Economic Interaction}

To address this need for rigorous evaluation, recent research has focused on benchmarking LLMs in specific negotiation and bargaining scenarios. Bianchi et al.\ introduced NegotiationArena, a framework for evaluating the negotiation capabilities of LLM-based agents \cite{bianchi2024well}. In this work, the authors implement three types of negotiation scenarios and show that, while LLMs can engage in deal-making, they are often prone to suboptimal strategic behaviors. However, the rapid improvement in LLMs' general reasoning capabilities and the continual release of new models have quickly made these results outdated. This motivates revisiting these experimental settings with frontier models, in order to determine whether superior reasoning capabilities mitigate these flaws or whether they persist despite scaling.

Complementing this perspective, Kwon et al.\ \cite{kwon2024llms} propose a strategy to measure the wide range of capabilities needed for effective negotiation, decomposing the task into three components: context understanding, coherent and strategically appropriate generation, and theory of mind. Although this work relies on analyzing static negotiation dialogue datasets rather than simulating full, dynamic interactions between agents, it remains a seminal contribution that highlights critical reasoning limitations of LLM-based systems.

Beyond capability benchmarking, recent research has shifted toward the practical and economic implications of these agents. Zhu et al.\ \cite{zhu2025automated} investigate the risks of fully automating negotiation in consumer markets. Their analysis characterizes agent-to-agent negotiation as an ``inherently imbalanced game,'' in which parties relying on less capable models face systematic economic disadvantages. Furthermore, they identify behavioral anomalies that translate into economic risk, such as ``constraint violation,'' where agents disregard user-imposed budget limits, and ``negotiation deadlock,'' where agents fail to reach mutually beneficial agreements despite overlapping interests. 

These findings underscore that the deployment of LLM-based agents in markets involves significant financial stakes. While these studies quantify the economic consequences of agentic flaws in specific consumer scenarios, our work complements this perspective by expanding the analysis to foundational bargaining games and frontier models. We show that these exploitative patterns are not isolated anomalies, but correlate with persistent cognitive biases that remain prevalent despite advances in model capabilities.

\subsection{Cognitive Biases and Game-Theoretic Perspectives}

Extensive research has shown that LLMs exhibit systematic cognitive biases that can compromise their decision-making and applicability in high-stakes scenarios. For instance, Cheung et al.\ \cite{cheung2025large} compare LLMs with human participants in realistic ethical dilemmas and find a strong ``omission bias'': a preference for inaction over action, even when acting could lead to better outcomes. Moreover, Lou et al.\ \cite{lou2024anchoring} demonstrate that LLMs are highly susceptible to anchoring, a cognitive bias in which the first piece of information encountered (the ``anchor'') disproportionately influences the final decision. Notably, the authors report that standard prompting techniques, such as Chain-of-Thought (CoT), are insufficient to mitigate this issue. Another prominent example is presented by Singh et al.\ \cite{singh2024large}, who provide empirical evidence of an overconfidence bias analogous to the Dunning--Kruger effect: LLMs tend to assign high confidence scores to their responses even when they are wrong. In fact, other work has pointed out that this bias can discourage the use of out-of-the-box LLMs as confidence estimators in classification tasks \cite{wen2024mitigating}.

The presence of such cognitive biases poses a fundamental challenge to the assumption of strategic rationality. According to classical game theory, rational agents driven by fixed incentives are expected to reach a stable Nash equilibrium \cite{nash1951non}. Furthermore, in many non-zero-sum negotiation and coordination settings with multiple equilibria, behavioral and experimental economics have shown that agents often coordinate on solutions that are socially prominent or perceived as fair, such as equitable splits in symmetric bargaining problems, even in the absence of communication \cite{schelling1960strategy,guth1982experimental}. In light of these biases, it remains unclear whether more capable LLMs move closer to classical equilibrium and focal-point behavior, or instead diverge into model-specific strategies. In this work, we directly address this question through controlled experiments in bargaining games.

\section{Methodology}

To empirically investigate the strategic behavior of frontier LLMs in negotiation settings, we adopt the NegotiationArena framework \cite{bianchi2024well}. This environment allows for the simulation of multi-turn, natural language negotiations between autonomous agents in three different negotiation games. In this section, we define the negotiation scenarios, the metrics used to evaluate strategic performance, and the models used in our simulations.

\subsection{Negotiation Scenarios}

\subsubsection{Seller and Buyer Scenario}
This scenario models a classic bilateral bargaining problem under incomplete information. In this setting, a Seller possesses an item with a private value $v_s$, representing the production cost, while a Buyer holds a private valuation $v_b$, representing their maximum willingness to pay. These values are expressed in arbitrary monetary units, denoted as ``ZUP''. To ensure the existence of a Zone of Possible Agreement (ZOPA), the game is initialized such that $v_b > v_s$. The agents engage in a multi-turn natural language dialogue to agree on a transaction price $P$, with the objective of maximizing their individual utility. The simulation terminates either when one agent accepts a proposed price or when a predefined maximum number of turns $T$ is reached. The seller always starts the conversation, representing a scenario in which the item is already for sale. Specifically, the strategic performance is measured by the agent's realized utility, defined as the payoff: $P - v_s$ for the Seller and $v_b - P$ for the Buyer.

\subsubsection{Multi-Turn Ultimatum Game}

Inspired by the classic Ultimatum game \cite{guth1982experimental}, this game pits two agents in a situation in which they have to split some resources. One agent (the Proposer) is initially endowed with a total resource quantity $C$, while the other agent (the Responder) holds the power to accept or reject the proposed division. The Proposer seeks to retain the maximum possible amount $C-x$, while the Responder aims to maximize their share $x$. If the Responder accepts a proposal, the resources are distributed according to the agreed-upon split. However, if the Responder rejects the final offer or if the negotiation exceeds the maximum turn limit $T$ without reaching an agreement, both agents receive a payoff of 0. The proposer initiates the conversation. The payoff for each agent is strictly defined as the final amount of resources secured upon agreement: $C-x$ for the Proposer and $x$ for the Responder.

\begin{figure}[!t]
    \centering
    \includegraphics[width=0.65\linewidth]{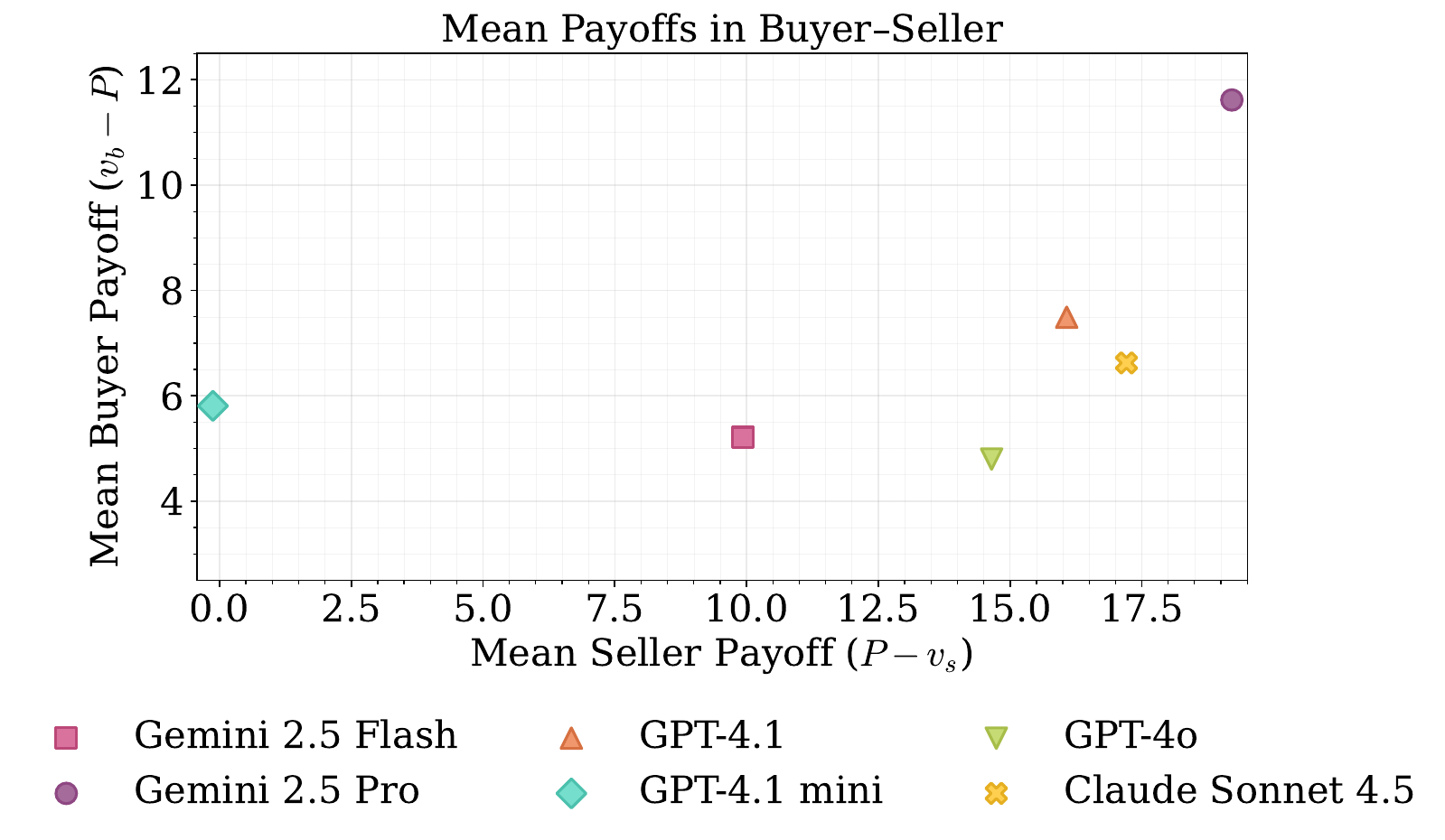}
    \caption{\textbf{Strategic Divergence in the Buyer-Seller Scenario.} The scatter plot illustrates the mean payoffs for Buyer ($v_b - P$) versus Seller ($P - v_s$) for each model. The results show a clear divergence in strategic behavior rather than convergence to a single equilibrium. \textbf{Gemini 2.5 Pro} (purple circle) achieves the highest combined utility, attaining higher payoffs as a buyer while maintaining strong seller performance. In contrast, \textbf{GPT-4.1 mini} (teal diamond) captures relatively little surplus in the seller role, 
residing in the lower-left region of the strategic landscape.}

    \label{fig:buy-sell-results}
\end{figure}

\subsubsection{Resource Exchange Game}

Each agent is initialized with a private endowment of resources (e.g., a specific quantity of distinct items $X$ and $Y$). Unlike the previous scenarios with strictly defined monetary value, the agents are assigned a broader objective: to ``maximize their total resources.'' This intentionally general goal allows for the emergence of diverse strategic behaviors; for instance, agents may decide to either diversify their resources or maximize only a single resource. Similar to the Seller and Buyer scenario, the interaction proceeds as a multi-turn conversation in natural language, in which agents propose an amount of resources to exchange for the counterpart's resources. The simulation terminates when a deal is accepted by any of the players or when a maximum number of turns $T$ is reached. Player 1 always initiates the conversation. The payoff is defined as the total net change in inventory across the set of item types $\mathcal{I}$, calculated as $R = \sum_{i \in \mathcal{I}} (q_{i}^{\text{final}} - q_{i}^{\text{initial}})$, where $q_{i}^{\text{final}}$ and $q_{i}^{\text{initial}}$ denote the quantity of item $i$ held by the agent at the end and start of the negotiation, respectively.

\subsection{Large Language Models}
\label{sec:methodology-llms}

To analyze the strategic performance of frontier models, we selected a diverse set of state-of-the-art models from different families available as of late 2025. Our selection criteria focused on models with demonstrated high performance in reasoning benchmarks, covering both ``Pro'' variants and ``Flash/Mini'' variants to assess the impact of model scaling on strategic behavior. Specifically, we evaluate the following models:

\begin{itemize} 
\item \textbf{Google DeepMind:} Gemini 2.5 Pro and Gemini 2.5 Flash. 
\item \textbf{OpenAI:} GPT-4.1, GPT-4o, and GPT-4.1 mini. 
\item \textbf{Anthropic:} Claude 4.5 Sonnet. 
\end{itemize}

All models were accessed via their respective official APIs. To ensure rigorous comparisons and minimize the variance introduced by stochastic decoding, we set the sampling temperature to 0.7 for all experiments. All models were prompted as in the original NegotiationArena implementation.

\section{Experiments and Results}

\begin{figure}[!t]
    \centering
    \includegraphics[width=0.6\linewidth]{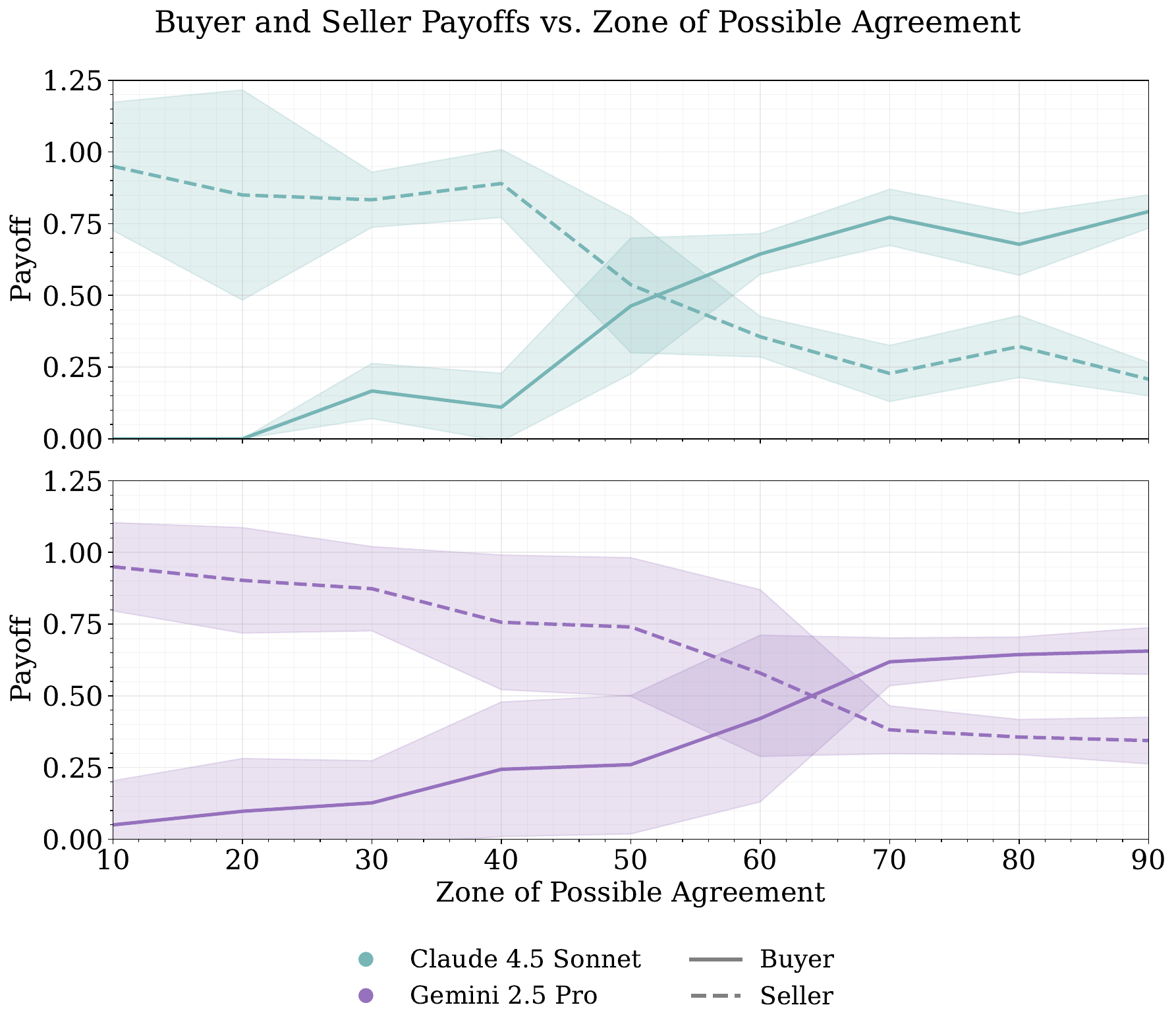}
    \caption{\textbf{Sensitivity to Negotiation Gap in the Buyer-Seller Scenario} The plots illustrate the evolution of mean Buyer (solid line) and Seller (dashed line) payoffs as the Zone of Possible Agreement (ZOPA) expands from 10 to 90. \textbf{(Top) Claude 4.5 Sonnet} exhibits a crossover point occurring near a gap of 50. \textbf{(Bottom) Gemini 2.5 Pro} displays a stronger seller bias, maintaining an advantage for a wider range of scenarios and shifting the crossover point to a larger gap ($\approx 65$). The shaded areas indicate payoff variance, highlighting the stability of the chosen strategy across episodes.}
    \label{fig:gap-analysis}
\end{figure}

\subsection{Strategic Divergence}

We first examine the general strategic positioning of the models in the Buyer-Seller scenario. To that end, we conducted a comprehensive evaluation in which each model played a total of 120 games per role, comprising 20 games against each of the 6 possible opponents detailed in Section \ref{sec:methodology-llms}. In each game, the seller holds a private valuation $v_{s}$ of 40 ZUP, while the buyer holds a private valuation $v_{b}$ of 60 ZUP, creating a Zone of Possible Agreement (ZOPA) of 20 ZUP. Fig. \ref{fig:buy-sell-results} illustrates the mean payoffs for each model across both roles. Contrary to the expectation of convergence toward a unified rational equilibrium, we observe a marked strategic divergence. The models do not cluster; instead, they occupy distinct strategic niches. For instance, Claude 4.5 Sonnet acts as a strong seller, achieving the second-highest score, yet clearly underperforms as a buyer.
Conversely, GPT-4.1 mini fails to maximize utility in either role, residing in the lower-left quadrant of the strategic landscape. Notably, Gemini 2.5 Pro emerges as the only model that tends to secure a higher payoff compared to its opponents in both roles. 
Interestingly, while general benchmarks rank GPT-4.1 and GPT-4.1 mini with very close scores (reflecting similar reasoning capabilities in static tasks), our dynamic negotiation environment reveals a substantial performance gap.

\begin{figure}[!t]
    \centering
    \includegraphics[width=0.65\linewidth]{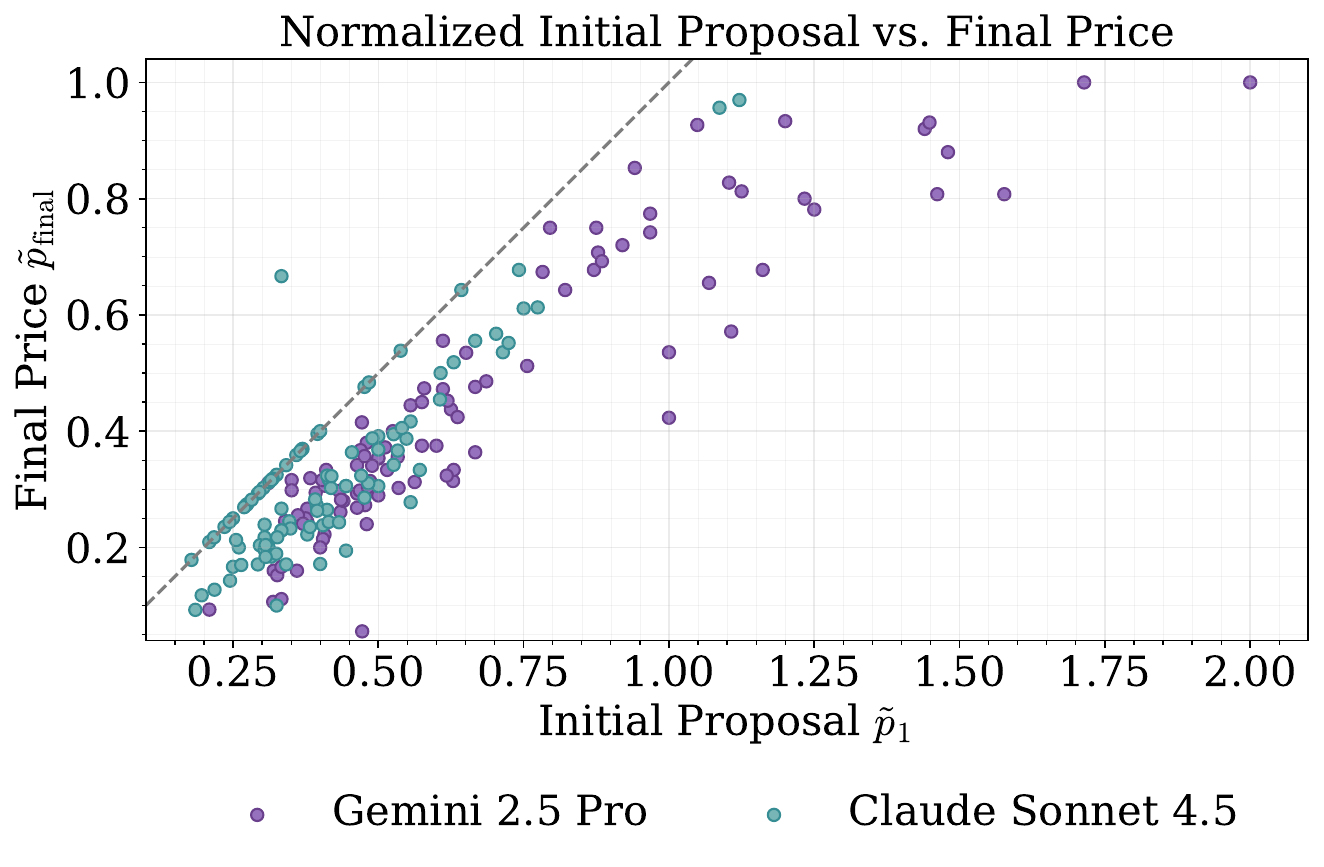}
    \caption{\textbf{Evidence of Numerical Anchoring Bias in Frontier Models.} Scatter plot of the normalized Final Price ($\tilde{p}_{\text{final}}$) versus the normalized Initial Proposal ($\tilde{p}_1$) in self-play simulations ($N=100$). The dashed diagonal line represents the identity function ($y=x$), corresponding to a scenario where the final outcome is perfectly determined by the initial anchor. The tight clustering of data points along this diagonal for both \textbf{Claude 4.5 Sonnet} ($\rho \approx 0.78$) and \textbf{Gemini 2.5 Pro} ($\rho \approx 0.91$) empirically demonstrates that superior reasoning capabilities do not eliminate susceptibility to anchoring heuristics.}
    \label{fig:numerical-anchoring}
\end{figure}

\begin{figure*}[!t]
    \centering
    \includegraphics[width=0.9\linewidth]{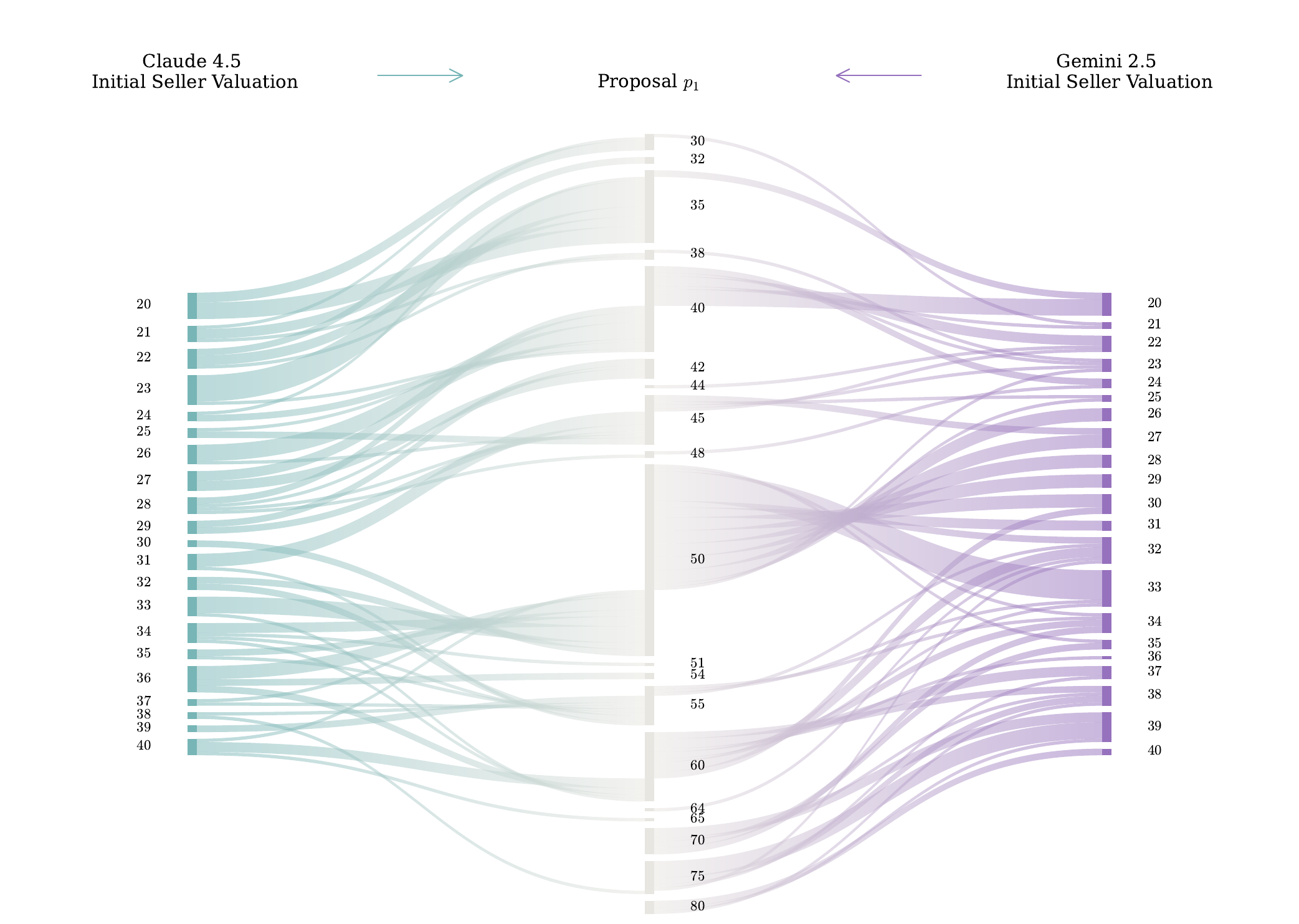} 
    \caption{\textbf{Visualizing Semantic Anchoring} The Sankey diagram maps the decision flow from the initial Seller Valuation ($v_s$, outer columns) to the generated Initial Proposal ($p_1$, center column) across 100 self-play episodes. \textbf{(Left)} Flows for Claude 4.5 Sonnet; \textbf{(Right)} Flows for Gemini 2.5 Pro. The convergence of diverse valuation inputs into a limited set of output bands (e.g., the heavy concentration on $p_1=50$) indicates the presence of a semantic anchoring tendency.}
    \label{fig:semantic-bias}
\end{figure*}

This divergence reflects fundamental differences in how models perceive and react to the incentive structure of the negotiation. To quantify this, we conducted a \textit{Gap Analysis}, varying the ZOPA size from 10 to 90 to observe how players' payoffs behave. For each scenario, we performed 20 simulations via self-play. Fig. \ref{fig:gap-analysis} reveals that each model possesses a unique ``strategic signature,'' defined by the specific threshold at which the balance of power shifts between the Seller and the Buyer. 
For instance, Claude 4.5 Sonnet (top) displays a crossover point occurring near a gap of 50. In contrast, Gemini 2.5 Pro (bottom) exhibits a distinct profile, maintaining a seller advantage for a wider range of scenarios and only ceding dominance at larger gaps ($\approx 65$). This disparity suggests that each model operates under a different internal valuation heuristic: Gemini 2.5 Pro appears to attribute a higher subjective value to the asset when acting as a seller, leading to a higher resistance threshold before conceding to the buyer's offers. Consequently, the point at which the bargaining power flips is not universal but is a function of the model's specific strategic behavior.

\subsection{Cognitive Biases}

Having established the divergence in strategic outcomes, we now investigate the cognitive mechanisms correlated with these behaviors. Specifically, we examine whether frontier models exhibit Anchoring Bias, a heuristic where the final decision is disproportionately influenced by the initial information received. While foundational studies using the NegotiationArena framework documented strong anchoring effects in earlier model generations \cite{bianchi2024well}, it remains unclear whether the substantial advancements in reasoning capabilities observed in modern frontier models have eliminated this bias, or if it persists despite scaling.

To answer this question, we replicated the original experimental setup proposed by Bianchi et al., simulating 100 games via self-play. For each episode, private valuations $v_{s}$ and $v_{b}$ were sampled from uniform distributions $U(20, 40)$ and $U(60, 80)$, respectively. We normalized both the initial proposal ($p_1$) and the final price ($P$) relative to the available surplus (ZOPA), defining the normalized price as $\tilde{p} = \frac{p - v_s}{v_b - v_s}$. Thus, $\tilde{p}=0$ corresponds to the Seller's valuation and $\tilde{p}=1$ to the Buyer's valuation.

Fig. \ref{fig:numerical-anchoring} presents the relationship between the normalized Initial Proposal ($\tilde{p}_1$) and the normalized Final Price ($\tilde{p}_{\text{final}}$). The results reveal a clear persistence of the bias. The data points tightly cluster along the diagonal, suggesting that the final settlement is heavily determined by the initial anchor. This observation is statistically corroborated by high Spearman rank correlations: $\rho \approx 0.78$ for Claude 4.5 Sonnet and a strong $\rho \approx 0.91$ for Gemini 2.5 Pro. To test the robustness of these findings, we replicated the anchoring experiments at temperatures of $0.0$ and $1.0$. Our results confirm that anchoring bias persists strongly across all regimes, counter-intuitively increasing or remaining saturated at higher temperatures. Specifically, Gemini 2.5 Pro remained consistently highly anchored, with correlations rising slightly from $\rho \approx 0.90$ at $T=0.0$ to $\rho \approx 0.92$ at $T=1.0$. In contrast, Claude 4.5 Sonnet showed a marked sensitivity to noise, where $\rho$ increased from $0.71$ at $T=0.0$ to $0.90$ at $T=1.0$.

Furthermore, we identify a more profound form of bias, semantic Anchoring. We carefully analyzed the decision pathways in the same batch of 100 self-play episodes to determine how the Seller translates the initial valuation ($v_s$) into the opening proposal ($p_1$). Fig. \ref{fig:semantic-bias} visualizes this internal flow using a Sankey diagram. Instead of the smooth, continuous distribution one might expect from a precise calculation of optimal proposals, the flows collapse into distinct price points (e.g., gravitating heavily towards 50 or other multiples of 5). This pattern suggests that the models do not compute an opening price solely as a function of the specific $v_s$, but instead tend to generate numbers aligned with common or salient magnitudes regardless of the underlying cost. This regularity may increase predictability in some cases, potentially leading to exploitable patterns.

\begin{figure}[!t]
    \centering
    
    \subfloat[\textbf{Multi-Turn Ultimatum Game.} Proposer Win Rates \textbf{(left)} and Average Payoffs \textbf{(right)}.]{
        \includegraphics[width=0.65\linewidth]{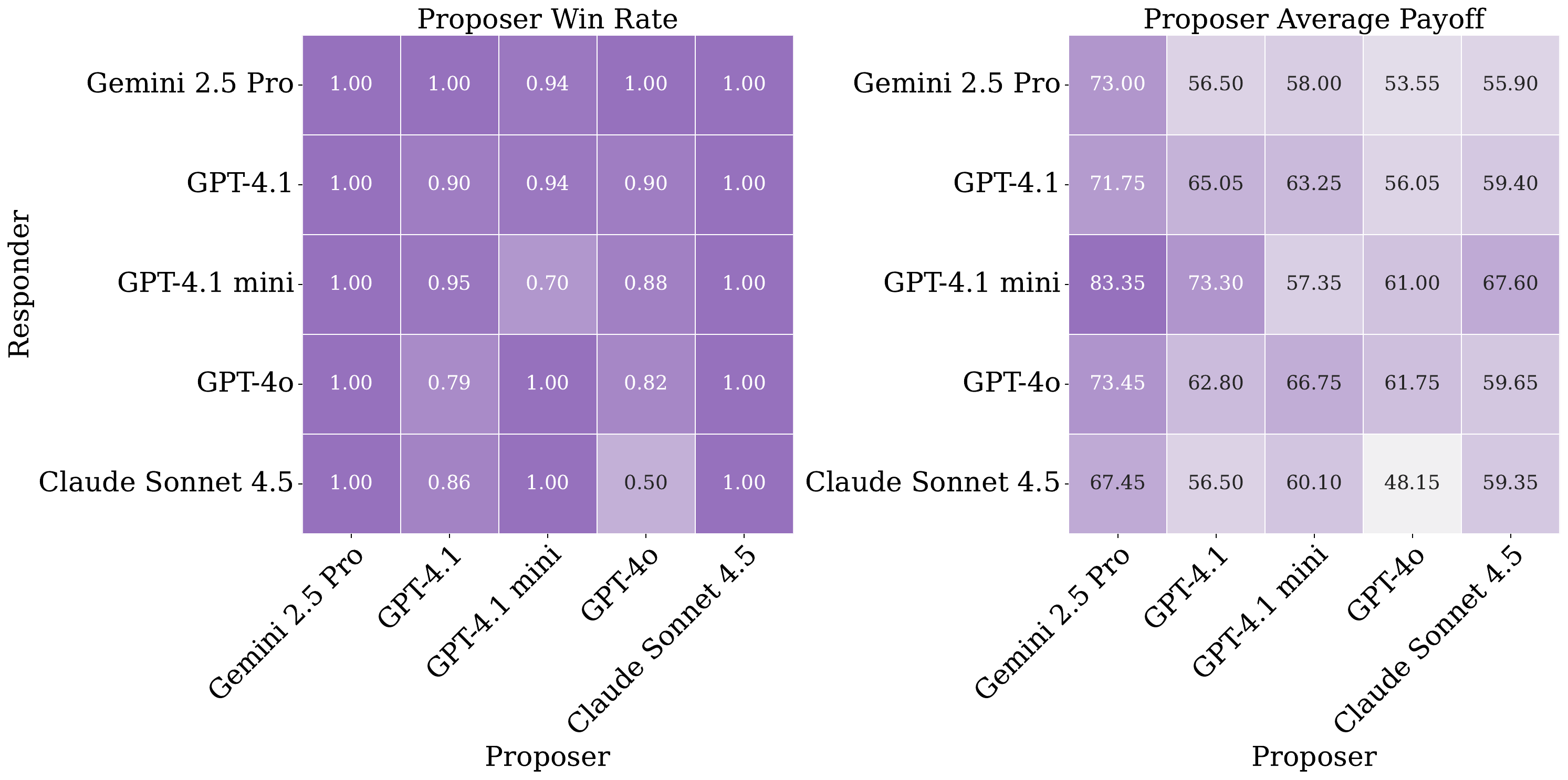}
        \label{fig:ultimatum-results}
    } 
    \\ 
    \subfloat[\textbf{Resource Exchange Game.} Player 2 Win Rates (left) and Average Payoffs (right).]{
        \includegraphics[width=0.65\linewidth]{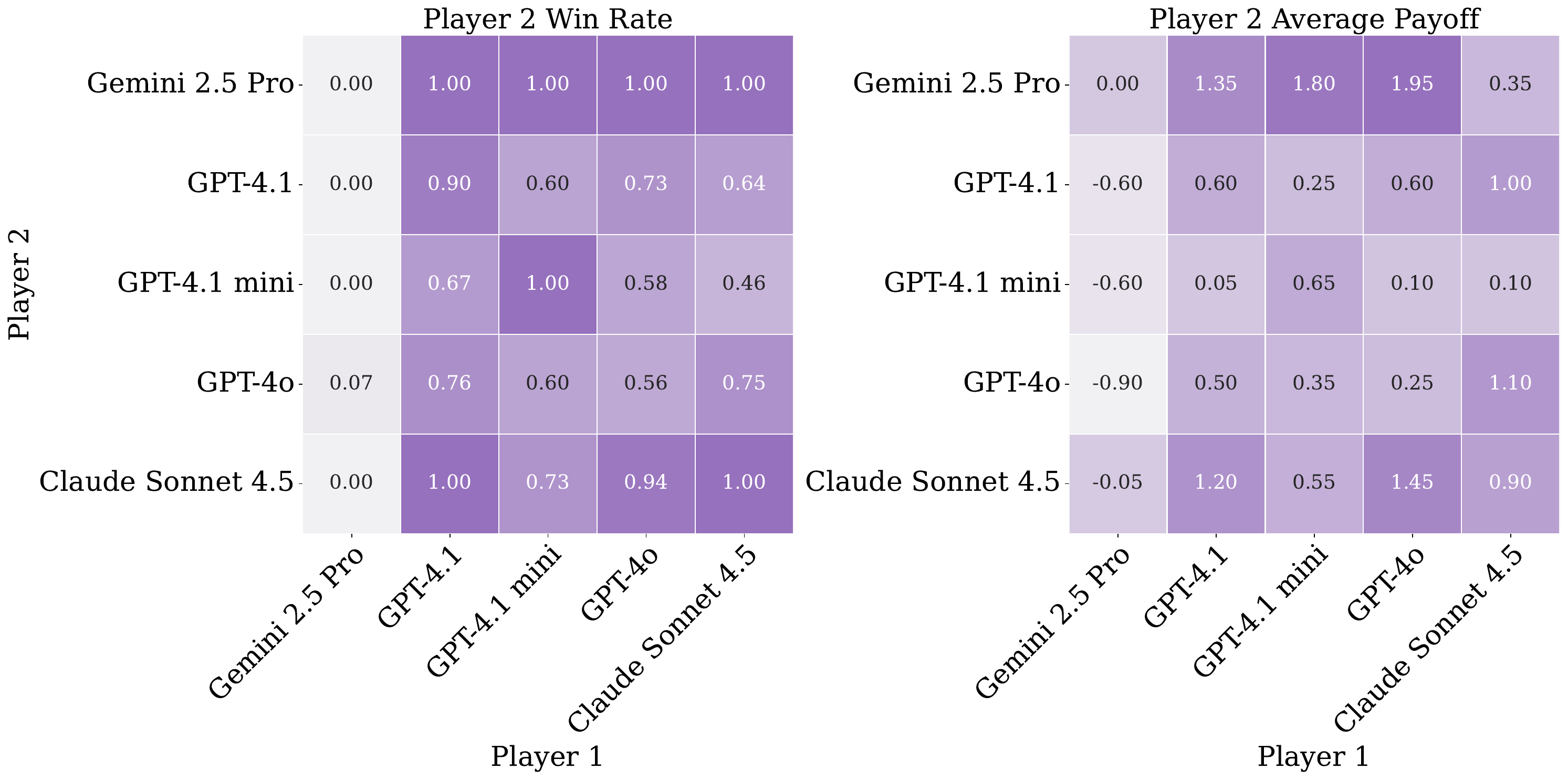}
        \label{fig:exchange-results} 
    }
    
   \caption{\textbf{Dominance and Asymmetry in Negotiation Outcomes.} The heatmaps display pairwise performance metrics across the two scenarios. The visual asymmetry across columns reveals a hierarchical landscape where advanced models (e.g., Gemini 2.5 Pro) tend to achieve higher payoffs than weaker counterparts (e.g., GPT-4.1 mini). Moreover, the results show a clear advantage for certain roles, creating asymmetric negotiation dynamics.}
    \label{fig:ultimatum-exchange-results} 
\end{figure}

\subsection{Dominance and Inequality}

Finally, we analyze the systemic consequences of these divergent strategies and persistent biases. Specifically, we investigate whether the strategies employed by frontier models lead to interactions in which some models consistently achieve lower payoffs than others. To that end, we leveraged the Multi-Turn Ultimatum and Resource Exchange games, replicating the original experimental setup proposed by Bianchi et al. \cite{bianchi2024well}. Fig. \ref{fig:ultimatum-exchange-results} presents the resulting payoff and win-rate matrices.

In the Multi-Turn Ultimatum Game (Fig. \ref{fig:ultimatum-results}), we observe that "stronger" models effectively leverage their position to achieve consistently higher payoffs from weaker counterparts. This is evident in the Proposer column associated with Gemini 2.5 Pro, which presents payoffs noticeably higher than those obtained by other proposers. Moreover, when Gemini 2.5 Pro acts as the Proposer against GPT-4.1 mini, it secures an average payoff of 83.35, indicating an asymmetric outcome that disadvantages the weaker model and reflecting the opponent's difficulty in negotiating more balanced splits. 
Similarly, the Resource Exchange Game (Fig. \ref{fig:exchange-results}) reveals comparable patterns of dominance. Gemini 2.5 Pro consistently outperforms its counterparts when acting as Player 1, creating a scenario resembling severe outcome asymmetry. Similar asymmetric patterns are also present in the Buyer and Seller scenario (Fig. \ref{fig:buy-sell-results}), where Gemini 2.5 Pro, once again,  tends to obtain larger surpluses than its counterparts, while GPT-4.1 mini consistently obtains the lowest payoffs in this setting.

\section{Conclusions and Future Work}

This work provides a comprehensive evaluation of frontier large language models in negotiation environments, challenging the assumption that scaling naturally leads models to converge toward a set of similar, rational, and unbiased strategies. We revisit prior analyses of LLM negotiation capabilities and extend them by evaluating a new generation of frontier models. Across all scenarios, our findings show that increased general reasoning abilities do not result in convergence toward a shared rational equilibrium. Instead, models consistently diverge into distinct, model-specific strategic behaviors, each exhibiting a characteristic strategic signature. Moreover, we find that these strategic signatures are shaped by persistent cognitive biases that remain evident even in frontier models.

Our results suggest that frontier models remain highly susceptible to both numeric and semantic anchoring. In our multi-turn bargaining simulations, initial proposals remained strongly predictive of final agreements, and models frequently gravitated toward salient or schematized values when generating opening offers. More critically, across all simulated scenarios, we observe clear dominance patterns in which stronger models consistently extract higher payoffs from weaker counterparts. These findings have significant implications for the deployment of AI agents in real high-stakes contexts, where systematic imbalances in strategic capability may lead to inequitable outcomes.

These findings indicate that developing rational negotiation systems based on LLMs will require a paradigm shift away from relying solely on model scaling. Accordingly, future work should explore mechanisms that mitigate the cognitive biases and strategic instabilities identified in this study. Furthermore, as these agents transition from closed sandbox environments to open economic settings, research should investigate defensive mechanisms that ensure fair and robust interactions among heterogeneous agents.

\bibliographystyle{spp-bst}
\bibliography{bibfile}

\end{document}